\documentclass{article}
\usepackage[utf8]{inputenc}

\usepackage{amsmath}
\usepackage{amssymb}

\title{Quantum Shape Kinematics}
\date{November 2015}

\author{Furkan Semih Dündar\footnote{email: furkan.dundar1@ogr.sakarya.edu.tr} $\,^{a}$ Barış Tamer Tonguç\footnote{email:btonguc@sakarya.edu.tr} $\,^{b}$\\ 
\footnotesize $^a$\textit{Physics and Mathematics Departments, Sakarya University, 54050, Sakarya, Turkey}\\
\footnotesize $^b$\textit{Physics Department, Sakarya University, 54050, Sakarya, Turkey}
}

\newcommand{\up}{\uparrow}
\newcommand{\down}{\downarrow}
\newcommand{\ket}[1]{\lvert #1 \rangle}

\newcommand{\complex}{\mathbb C}

\begin{document}

\maketitle

\begin{abstract}
Shape dynamics is a theory first proposed by Julian Barbour which states that physics happen uniquely in the reduced configuration space of a theory. So far, studies in the area have focused on gravitational systems. Here we first contemplate on the implications of this idea on quantum mechanics. We summarize the idea of shape dynamics and then give physical configurations of  multi qubit systems. It is observed that inclusion of the spacetime curvature introduces a new qubit degree of freedom. This is a novel effect of general relativity on a quantum system. Our aim in the grand picture, is to initiate a research program translating classical shape dynamics to quantum realm.
\end{abstract}

\section{Introduction}

Mach's principle may refer to many ideas. Hermann Bondi and Joseph Samuel list eleven distinct versions of the Mach's principle that can be found in the literature \cite{bondimach}. We adopt the Julian Barbour's definition \cite{jbmach}.

First of all there is the configuration space of a theory. By removing the gauge degree of freedoms we obtain the reduced configuration space of a theory where all degrees of freedom are physical. This space is called the \emph{shape space}. The Mach's principle states that \cite{jbmach} a point and a direction or a tangent vector in shape space determine the evolution of the system uniquely.

According to classical shape dynamics, in the Newtonian $N$-body problem the physical configurations are obtained when the rotation and scale degrees of freedom are removed \cite{sd-intro}. Hence for one or two particles in an empty universe there is no degree of freedom. Non-trivial Shape dynamics apply to the cases of three or more particles. For an introduction to shape dynamics, reader may refer to \cite{sd-intro} and \cite{sd-tutor}.

So far the studies on shape dynamics focused on gravitational \cite{thin-shell} and classical aspects such as the arrow of time \cite{arrow-of-time}. Beginning with this paper, we would like to initiate a research endeavour that investigates consequences of classical shape dynamics in quantum phenomena.

\section{Shapes of Qubits}

We discuss quantum shape kinematics of multi qubit systems beginning with the cases of a single and a double qubit systems. In this section qubits do not occupy positions in spacetime. Hence they have only internal degrees of freedom.

\subsection{Single qubit system}
We consider there exists only one qubit in the Minkowski spacetime and nothing else. The quantum state of the particle can be written as:

\begin{equation}
\ket \psi = \alpha \ket \up + \beta \ket \down, \quad \exists \alpha,\beta \in \complex
\end{equation}

However by rotating the coordinates and multiplying with a complex number we can always map $\ket \psi \mapsto \ket \up$. Therefore we conclude that for one qubit there is no physical degree of freedom apart from its mere existence.

\subsection{Double qubit system}

Here we suppose there are two qubits in the universe. The basis vectors are $\ket{\up\up}, \ket{\down\down}, \ket{\up\down}$ and $\ket{\down\up}$. Because all we have is the angles between spins, the first two correspond to parallel spin case, $\ket{\text{parallel}}$, and the last two correspond to anti-parallel spin case, $\ket{\text{anti-parallel}}$. We can always rotate the state of first qubit into $\ket \up$. Hence the direction of the first spin is used to fix a direction in space. The physical basis vectors are $\ket{\up\up}$ and $\ket{\up\down}$.

\subsection{Multiple qubit systems}

In this part we suppose there are $N$ qubits in the universe. The basis vectors of the system are $\ket{a_1}\otimes\ket{a_2}\otimes \cdots\otimes \ket{a_N}$ where $\ket{a_i}$ for $1 \leq i \leq N$ can be $\ket \up$ or $\ket \down$. Whatever the value of $\ket{a_1}$ we can always rotate it to $\ket\up$. Hence we reduce one degree of freedom. We call the first qubit as \emph{the reference qubit}. $N$ qubit system has the degrees of freedom of $N-1$ qubit system. This is true for all $N \geq 1$.

\section{Inclusion of spacetime curvature}

In this section we put our qubit systems in different locations in a curved spacetime manifold. We suppose there are two qubit systems each containing $N_1$ and $N_2$ qubits located at points $p_1$ and $p_2$. If there is no interaction between the two systems, we have $N_1-1$ qubit degrees of freedom for the first system, whereas the second system has $N_2-1$ qubit degrees of freedom. This is a local assessment. However when we think globally, we see that there must be $N_1+N_2-1$ degrees of freedom. The difference between local and global viewpoints results in a new qubit degree of freedom: the reference qubit of the second system becomes a dynamical entity. What is more, the ket expansion of the second system should be done through the direction of the parallel transported form of the direction of the first reference qubit. This dependence necessarily involves the curvature of spacetime manifold. Hence we observe a novel effect of gravity on a quantum system.

Let $n^\mu$ be the direction of the first reference qubit. In order to find the direction of the second reference qubit we need to parallel transport $n^\mu$. If $x^\mu(\lambda)$ is a geodesic curve connecting the points $p_1$ and $p_2$ and $\lambda$ an affine parameter, the parallel transport equation is given as follows \cite{carroll-gr-book}:

\begin{equation}
\frac{dx^\nu}{d\lambda}\nabla_\nu n^\mu = \frac{dn^\mu}{d\lambda} + \Gamma_{\alpha \beta}^\mu \frac{dx^\alpha}{d\lambda}n^\beta = 0
\end{equation}

Parallel transport does not change the length $n^\mu n_\mu$ of the direction vector which we take to be of unit norm. However this process may rotate the vector. Rotations in 3D space has two parameters which can be represented on a unit sphere. On the other hand, we observe that the state of a qubit can be represented on a two sphere as well. Here we have obtained a direct correspondence. The new degree of freedom is a direct manifestation of the spacetime curvature. Therefore in total, the global quantum system has $N_1+N_2$ many degrees of freedom. This is an example where global thinking introduces new degrees of freedom. We therefore conclude that reductionism is incapable of explaining quantum phenomana in curved spacetimes.

\section{Possible Objections}
\subsection{Interactions}

One may object to this classification with the counter example of interacting two qubits via the Hamiltonian $-\gamma \vec{\sigma}_1 \odot \vec{\sigma}_2$. The eigenstates of the systems are $\ket{\up\up},\ket{\down\down}$ and $\ket{\up\down}+\ket{\down\up}$ with energy eigenvalue $-\gamma$, whereas the state $\ket{\up\down}-\ket{\down\up}$ has energy $3\gamma$. The seemingly paradoxical point is that the states  $\ket{\up\down}-\ket{\down\up}$ and $\ket{\up\down}+\ket{\down\up}$ have different energies though they should correspond to the same  $\ket{\text{anti-parallel}}$ state vector.

We need to note that in order for two qubits to \emph{interact}, we need to introduce additional structure. In quantum shape dynamics, time evolution should be reached by considering the whole state of the universe. It will be a holistic theory.

For simplicity suppose that the interaction between the qubits introduce another qubit. All in all the system become a triple qubit system. Here the physical states are such that the additional qubit has always the state $\ket\up$. Basis vectors are $\ket\up \otimes \ket{a} \otimes \ket b$ where $\ket a, \ket b$ can be $\ket\up$ or $\ket\down$. It is now that the states $\ket{\up\up\down}-\ket{\up\down\up}$ and $\ket{\up\up\down}+\ket{\up\down\up}$ are physically different and there is no paradox for them having different energy values. The problem is solved once we take into account the whole system.

\subsection{Choice of the reference qubit}

The choice of the reference qubit, the qubit whose state is fixed, is arbitrary. By definition it has no dynamics, though it may evolve relative to subsystems.

\subsection{Parallel transport of $n^\mu$}

In a curved spacetime the result of the parallel transport of a vector depends on the curve chosen between two points. Among the multitude of possible curves, we chose the geodesic curve. This choice gave us a unique curve. This selection is made so that in the flat spacetime one parallel transports $n^\mu$ on a line.

\section{Conclusion}

In this study, we removed the rotation gauge degrees of freedoms from multi qubits system and what has remained is the quantum shapes of the system. We observed that one qubit can be put in a fixed state by rotating the space, hence it has no degree of freedom. In the general case of an $N$ qubit system, the degrees of freedom turned out to have the degrees of freedom of $N-1$ qubit system.

Secondly, we though of two qubit systems at different locations in spacetime. We have seen that by thinking globally, the reference qubit of the second system has become a dynamical entity. Moreover, spacetime curvature between the two states introduced another new qubit degree of freedom. This is a novel effect of spacetime curvature on quantum systems.

Future works may expand our discussion by including quantum shape kinematics of molecules or higher spin systems. Quantum shape dynamics solution for the Hydrogen atom would be an interesting example to see. It seems that there will be a difference on the formation of a single hydrogen atom in the universe compared to formation of $N \gg 1$ hydrogen atoms.

\section*{Acknowledgements}

F.S.D. would like to thank Julian Barbour for useful discussions.

\bibliographystyle{plain}
\bibliography{references}
\end{document}